\documentclass[prl,superscriptaddress,unsortedaddress,twocolumn,showpacs,floatfix]{revtex4}

\usepackage{epsfig}
\usepackage{dcolumn}
\usepackage{amsmath}

\newcommand{\ppzkin}{\mbox{$k_{+-0} $}}

\newcommand{\lzero}{\mbox{$\lambda_0$}}
\newcommand{\lplusp}{\mbox{$\lambda'_+$}}

\newcommand{\lpluspp}{\mbox{$\lambda''_+$}}

\newcommand{\deltae}{\mbox{$\delta_K^e$}}
\newcommand{\deltam}{\mbox{$\delta_K^\mu$}}

\newcommand{\fplust}{\mbox{${f}_+(t)$}}

\newcommand{\fzerot}{\mbox{${f}_0(t)$}}

\newcommand{\vus}{\mbox{$V_{us}$}}

\newcommand{\KLpienu}{\mbox{$K_{L}\to\pi^{\pm}e^{\mp}\nu$}}

\newcommand{\KLpimunu}{\mbox{$K_{L}\to\pi^{\pm}{\mu}^{\mp}\nu$}}

\newcommand{\Rzz}{\mbox{$\Gamma(K_{L}\rightarrow \pi^0\pi^0)$}}
\newcommand{\Rzzz}{\mbox{$\Gamma(K_{L}\rightarrow \pi^0\pi^0\pi^0)$}}

\newcommand{\Rpm}{\mbox{$\Gamma(K_{L}\rightarrow \pi^+\pi^-)$}}
\newcommand{\Rpienu}{\mbox{$\Gamma(\KLpienu)$}}
\newcommand{\Rpimunu}{\mbox{$\Gamma(\KLpimunu)$}}
\newcommand{\Rppp}{\mbox{$\Gamma(K_{L}\rightarrow \pi^+\pi^-\pi^0)$}}

\newcommand{\Gzz}{\mbox{$\Gamma_{00}$}}
\newcommand{\Gzzz}{\mbox{$\Gamma_{000}$}}
\newcommand{\Gpm}{\mbox{$\Gamma_{+-}$}}
\newcommand{\Gpienu}{\mbox{$\Gamma_{Ke3}$}}
\newcommand{\Gpimunu}{\mbox{$\Gamma_{K\mu3}$}}
\newcommand{\Gppp}{\mbox{$\Gamma_{+-0}$}}

 \newcommand{\beq}{\begin{equation}}
 \newcommand{\eeq}{\end{equation}}
 \newcommand{\bqa}{\begin{eqnarray}}
 \newcommand{\eqa}{\end{eqnarray}}

\newcommand{\KL}{\mbox{$K_{L}$}}
\newcommand{\KS}{\mbox{$K_{S}$}}

\def\epe{\epsilon'\!/\epsilon}

\newcommand{\etapm}{\mbox{$\eta_{+-}$}}

\def\kpi0{K_{L}\to 3\pi^0}
\def\ke3{K_{L}\to\pi^{\pm}e^{\mp}\nu}
\def\k2pi{K_{L} \to \pi^+\pi^-}

\newcommand{\KLpm}{\mbox{$K_{L}\rightarrow\pi^{+}\pi^{-}$}}

\newcommand{\KLzz}{\mbox{$K_{L}\rightarrow\pi^{0}\pi^{0}$}}

\newcommand{\KLzzz}{\mbox{$K_{L}\rightarrow \pi^{0}\pi^{0}\pi^{0}$}}
\newcommand{\KLpmz}{\mbox{$K_{L}\rightarrow \pi^{+}\pi^{-}\pi^{0}$}}

\def\RPMNvalue{0.6640}
\def\RPMNerrstat{0.0014}
\def\RPMNerrsyst{0.0022}

\def\RZZZvalue{0.4782}
\def\RZZZerrstat{0.0014}
\def\RZZZerrsyst{0.0053}

\def\RPPPvalue{0.3078}
\def\RPPPerrstat{0.0005}
\def\RPPPerrsyst{0.0017}

\def\RPPvalue{ 4.856}
\def\RPPerrstat{ 0.017}
\def\RPPerrsyst{ 0.023}

\def\RNEUTvalue{ 4.446}
\def\RNEUTerrstat{ 0.016}
\def\RNEUTerrsyst{ 0.019}

\def\BKEvalue{0.4067}

\def\BKEerrtot{0.0011}
\def\WKEvalue{0.7897}

\def\WKEerrtot{0.0065}
  
\def\BKMvalue{0.2701}

\def\BKMerrtot{0.0009}
\def\WKMvalue{0.5244}

\def\WKMerrtot{0.0044}
  
\def\BZZZvalue{0.1945}

\def\BZZZerrtot{0.0018}
\def\WZZZvalue{0.3777}

\def\WZZZerrtot{0.0045}
  
\def\BPMZvalue{0.1252}

\def\BPMZerrtot{0.0007}
\def\WPMZvalue{0.2431}

\def\WPMZerrtot{0.0023}
  
\def\BPMvalue{ 1.975}

\def\BPMerrtot{ 0.012}
\def\WPMvalue{ 3.835}

\def\WPMerrtot{ 0.038}
  
\def\BZZvalue{ 0.865}

\def\BZZerrtot{ 0.010}
\def\WZZvalue{ 1.679}

\def\WZZerrtot{ 0.024}
  
\def\vusfave {0.2165}
\def\vuske   {0.2253}
\def\vuskm   {0.2250}
\def\vusave  {0.2252}
\def\vusavein{0.0008}
\def\vusaveex{0.0021}

\def\vusfavee{0.0012}
\def\vuskee  {0.0023}
\def\vuskme  {0.0023}
\def\vusavee {0.0022}

\def\deltauni{0.0018}
\def\unite   {0.0019}
  
\def\etapmv  { 2.228}
\def\etapme  { 0.010}
\def\etapmint{ 0.005}
\def\etapmext{ 0.009}

\def\lplabv{20.64}
\def\lplbbv{3.20}
\def\lzbv{13.72}

\def\lplabe{1.75}
\def\lplbbe{0.69}
\def\lzbe{1.31}

\def\ikebv{0.15350}
\def\ikmbv{0.10165}

\def\ikrbv{0.6622}

\def\LeptUni{0.9969}
\def\LeptUniE{0.0048}
\def\DeltaRat{1.0058}
\def\DeltaRatE{0.0010}
\def\PDGLeptUni{1.0270}
\def\PDGLeptUniE{0.0182}
\def\ikrbetot{0.0018}
\def\ikebetot{0.00105}
\def\ikmbetot{0.00080}

\begin{document}

\title{A Determination of the CKM Parameter $|\vus|$
}

\newcommand{\UAz}{University of Arizona, Tucson, Arizona 85721}
\newcommand{\UCLA}{University of California at Los Angeles, Los Angeles,
                    California 90095} 
\newcommand{\UCSD}{University of California at San Diego, La Jolla,
                   California 92093} 
\newcommand{\EFI}{The Enrico Fermi Institute, The University of Chicago, 
                  Chicago, Illinois 60637}
\newcommand{\UB}{University of Colorado, Boulder, Colorado 80309}
\newcommand{\ELM}{Elmhurst College, Elmhurst, Illinois 60126}
\newcommand{\FNAL}{Fermi National Accelerator Laboratory, 
                   Batavia, Illinois 60510}
\newcommand{\Osaka}{Osaka University, Toyonaka, Osaka 560-0043 Japan} 
\newcommand{\Rice}{Rice University, Houston, Texas 77005}
\newcommand{\UVa}{The Department of Physics and Institute of Nuclear and 
                  Particle Physics, University of Virginia, 
                  Charlottesville, Virginia 22901}
\newcommand{\UW}{University of Wisconsin, Madison, Wisconsin 53706}

\affiliation{\UAz}
\affiliation{\UCLA}
\affiliation{\UCSD}
\affiliation{\EFI}
\affiliation{\UB}
\affiliation{\ELM}
\affiliation{\FNAL}
\affiliation{\Osaka}
\affiliation{\Rice}
\affiliation{\UVa}
\affiliation{\UW}

\author{T.~Alexopoulos}   \affiliation{\UW}
\author{M.~Arenton}       \affiliation{\UVa}
\author{R.F.~Barbosa}     \altaffiliation[Permanent address: ]
   {University of S\~{a}o Paulo, S\~{a}o Paulo, Brazil}\affiliation{\FNAL}
\author{A.R.~Barker}      \altaffiliation[Deceased.]{ } \affiliation{\UB}
\author{L.~Bellantoni}    \affiliation{\FNAL}
\author{A.~Bellavance}    \affiliation{\Rice}
\author{E.~Blucher}       \affiliation{\EFI}
\author{G.J.~Bock}        \affiliation{\FNAL}
\author{E.~Cheu}          \affiliation{\UAz}
\author{S.~Childress}     \affiliation{\FNAL}
\author{R.~Coleman}       \affiliation{\FNAL}
\author{M.D.~Corcoran}    \affiliation{\Rice}
\author{B.~Cox}           \affiliation{\UVa}
\author{A.R.~Erwin}       \affiliation{\UW}
\author{R.~Ford}          \affiliation{\FNAL}
\author{A.~Glazov}        \affiliation{\EFI}
\author{A.~Golossanov}    \affiliation{\UVa}
\author{J.~Graham}        \affiliation{\EFI}   
\author{J.~Hamm}          \affiliation{\UAz}
 
\author{K.~Hanagaki}      \affiliation{\Osaka}
\author{Y.B.~Hsiung}      \affiliation{\FNAL}
\author{H.~Huang}         \affiliation{\UB}
\author{V.~Jejer}         \affiliation{\UVa}  
\author{D.A.~Jensen}      \affiliation{\FNAL}
\author{R.~Kessler}       \affiliation{\EFI}
\author{H.G.E.~Kobrak}    \affiliation{\UCSD}
\author{K.~Kotera}        \affiliation{\Osaka}
\author{J.~LaDue}         \affiliation{\UB}
  
\author{A.~Ledovskoy}     \affiliation{\UVa}
\author{P.L.~McBride}     \affiliation{\FNAL}

\author{E.~Monnier}
   \altaffiliation[Permanent address: ]{C.P.P. 
    Marseille/C.N.R.S., France}\affiliation{\EFI}
\author{H.~Nguyen}       \affiliation{\FNAL}
\author{R.~Niclasen}     \affiliation{\UB} 
\author{V.~Prasad}       \affiliation{\EFI}
\author{X.R.~Qi}         \affiliation{\FNAL}
\author{E.J.~Ramberg}    \affiliation{\FNAL}
\author{R.E.~Ray}        \affiliation{\FNAL}
\author{M.~Ronquest}	 \affiliation{\UVa}
\author{E. Santos}       \altaffiliation[Permanent address: ]
      {University of S\~{a}o Paulo, S\~{a}o Paulo, Brazil}\affiliation{\FNAL}
\author{P.~Shanahan}     \affiliation{\FNAL}
\author{J.~Shields}      \affiliation{\UVa}
\author{W.~Slater}       \affiliation{\UCLA}
\author{D.~Smith}	 \affiliation{\UVa}
\author{N.~Solomey}      \affiliation{\EFI}
\author{E.C.~Swallow}    \affiliation{\EFI}\affiliation{\ELM}
\author{P.A.~Toale}      \affiliation{\UB}
\author{R.~Tschirhart}   \affiliation{\FNAL}
\author{Y.W.~Wah}        \affiliation{\EFI}
\author{J.~Wang}         \affiliation{\UAz}
\author{H.B.~White}      \affiliation{\FNAL}
\author{J.~Whitmore}     \affiliation{\FNAL}
\author{M.~Wilking}      \affiliation{\UB}
\author{B.~Winstein}     \affiliation{\EFI}
\author{R.~Winston}      \affiliation{\EFI}
\author{E.T.~Worcester}  \affiliation{\EFI}
\author{T.~Yamanaka}     \affiliation{\Osaka}
\author{E.~D.~Zimmerman} \affiliation{\UB}

\collaboration{The KTeV Collaboration}

\begin{abstract}
  We present a determination of the CKM parameter $|V_{us}|$ based
  on new measurements of the six largest $K_L$ branching fractions
  and semileptonic form factors by the KTeV (E832) experiment at
  Fermilab.  We find
   $|\vus| = \vusave \pm \vusavein_{\rm KTeV} \pm \vusaveex_{\rm ext}$,
   where the errors are from KTeV measurements and from external sources.   
   We also use the
   measured branching fractions to determine the CP violation
  parameter $|\etapm| = (\etapmv \pm \etapmint_{\rm KTeV} 
  \pm \etapmext_{\rm ext})\times 10^{-3} $.
  
\end{abstract}


\maketitle


The Cabibbo-Kobayashi-Maskawa (CKM) matrix~\cite{cabibbo,km} describes the 
charged current
couplings of the $u$, $c$, and $t$ quarks to the $d$, $s$, and $b$ quarks.
The first row of this matrix provides the 
most stringent test of the unitarity 
of the matrix. Current measurements~\cite{pdg02} deviate from unitarity
at the 2.2 sigma level: 
$1-(|V_{ud}|^2+|V_{us}|^2+|V_{ub}|^2)= 0.0043 \pm 0.0019$.  
$|V_{us}|$, which contributes an uncertainty of 0.0010 to this unitarity test, 
has been determined from charged and neutral kaon semileptonic 
decay rates. 
This determination is based on 
the partial width for semileptonic $K$ decay, $\Gamma_{K\ell 3}$:
\begin{equation}
\Gamma_{K\ell 3}=\frac{\textstyle G_F^2
M_K^5}{\textstyle 192\pi^3} S_{EW} (1+\delta^\ell_K)C^2
\left|\vus\right|^2f^2_+(0) I^\ell_K, 
\label{eq:vus}
\end{equation}
where $\ell$ refers to either $e$ or $\mu$,
$G_F $ is the Fermi constant,
$M_K$ is the kaon mass,
$S_{EW}$ is the  short-distance radiative correction,
$\delta^\ell_K$ is the mode-dependent long-distance radiative
correction, $f_+(0)$ is the calculated 
form factor at zero momentum transfer for the 
$\ell \nu$ system, and $I^\ell_K$ is the phase-space integral, which depends
on measured semileptonic form factors. $C^2$ is 1 (1/2) for neutral 
(charged) kaon
decays.
The current PDG determination of $|\vus|$ is based only on 
$K \to \pi e \nu$ decays; $K \to \pi \mu \nu$ decays have not been used
because of large uncertainties in $I^{\mu}_K$.

In this Letter, we present a determination of $|V_{us}|$ by the KTeV
(E832) experiment at Fermilab based on measurements of the
$\KLpienu$ and $\KLpimunu$ partial widths  and form 
factors.  These measurements are described in
detail elsewhere~\cite{ktev_kbr,ktev_kl3ff}; 
a brief summary is given here.
Our $|V_{us}|$
determination also makes use of a new treatment of
radiative corrections~\cite{Troy}. 

To determine the $\KLpienu$ and \KLpimunu\ partial widths,
we measure the following five ratios:
\begin{eqnarray}
\Gpimunu/\Gpienu & \equiv & \Rpimunu/ \Rpienu\  \\
\Gppp/\Gpienu  & \equiv & \Rppp/ \Rpienu\ \ \ \\ 
\Gzzz/\Gpienu  & \equiv & \Rzzz/ \Rpienu\  \\
\Gpm/\Gpienu  & \equiv & \Rpm/ \Rpienu\  \\
\Gzz/\Gzzz  & \equiv & \Rzz/ \Rzzz, 
\end{eqnarray}
where internal bremsstrahlung contributions are included for all
decay modes with charged particles.
Since the six decay modes listed above account for more than
99.9\% of the total decay rate, the five partial width ratios may be 
converted into measurements of
the branching fractions for the six decay modes. 
The $K_L$ lifetime
is then used to convert these branching fractions into partial widths.
The branching fraction measurements
also can be used to determine the CP violation parameter
$|\etapm|^2 \equiv \Gamma(\KL \to \pi^+\pi^-)/
\Gamma(\KS \to \pi^+\pi^-)$.

The semileptonic form factors 
describe the distribution of $t$, the momentum
transfer to the $\ell\nu$ system.  This $t$ 
dependence increases the decay phase space 
integrals, $I^{e}_K$ and $I^{\mu}_K$, by about 10\%.  
We use the following parametrization for the
two independent semileptonic form factors:
\begin{equation}
\begin{array}{l}
\fplust = f_+(0)\left(1 + \lplusp \frac{\textstyle t}{\textstyle M^2_\pi} +
\frac{\textstyle 1}{\textstyle 2} \lpluspp 
 \frac{\textstyle t^2}{\textstyle M^4_\pi}\right) \\
\fzerot = f_+(0)\left(1 + \lzero \frac{\textstyle t}{\textstyle
M^2_\pi}\right),  \\
\end{array}
\end{equation}
where $f_+(0)$ is obtained from theory, and we measure \lplusp, \lpluspp, and
\lzero.

In principle, the form factors can be measured directly from the $t$ 
distribution.  The undetected neutrino and unknown
kaon momentum, however, result in a
twofold ambiguity in the reconstructed value of $t$. To avoid 
systematic uncertainties associated with this ambiguity,
we use a technique based only 
on components of particle momenta measured
transverse to the kaon momentum~\cite{ktev_kl3ff}.

\begin{figure}
\centering
\psfig{figure=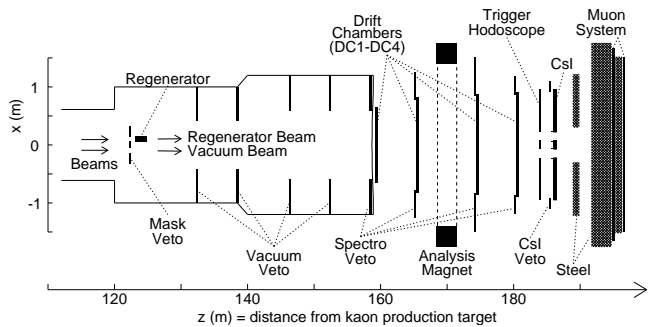,width=8.5cm}
\caption{The KTeV (E832) detector.}
\label{fig:detector}
\end{figure}

The KTeV experiment (Fig.~\ref{fig:detector})
and associated event reconstruction techniques have 
been described in detail elsewhere~\cite{Alavi-Harati:2002ye}. 
An 800 GeV/c proton beam striking a BeO target is used to produce two
almost parallel neutral beams. The regenerator beam,
which includes $K_S$, is
not used in this analysis; the vacuum beam
provides $K_L$ decays used for these measurements.  
A large vacuum decay region surrounded by
photon veto detectors extends to 159~m from the primary target. Following
the vacuum region is a drift chamber
spectrometer, trigger hodoscope, pure CsI electromagnetic calorimeter,
and a muon system consisting
of scintillator hodoscopes
behind 4 m and 5 m of steel.
The analyses presented in this Letter make use of the detector calibration
and Monte Carlo simulation from 
the KTeV $\epe$ analysis~\cite{Alavi-Harati:2002ye}.

Simple event reconstruction and selection may be
used to distinguish different kaon decay modes from each other, and to reduce 
background to a negligible level for all decay modes.  
The reconstruction of charged decay modes 
(\KLpienu, \KLpimunu, \KLpmz, and \KLpm)
begins with the identification of two oppositely charged tracks
coming from a single vertex. Note that for the $\KLpmz$ decay, we choose
not to reconstruct the $\pi^0 \to \gamma\gamma$ decay to reduce the acceptance
uncertainty in the $\Gppp/ \Gpienu$ ratio.

The  charged decay modes are separated  from each other
on the basis of particle identification and kinematic requirements.
To select $\KLpienu$ decays, the electron is identified using
the calorimeter energy measurement ($E$), 
combined with the spectrometer momentum ($p$).
\KLpm\ is separated from other two-track decays based on the 
two-track invariant mass,
$m_{\pi\pi}$, and the square of the two-track momentum transverse 
to the $K_L$ direction, $p^2_t$.
To isolate \KLpmz\ we use an additional variable, 
\ppzkin, described in \cite{ktev_kbr}.
Note that the \KLpimunu\ background to each of these decay 
modes is suppressed by rejecting events with hits 
in the muon system.
Two-track events 
that are not identified as \KLpienu, \KLpmz, or \KLpm,
are selected as \KLpimunu\ candidates.
To reduce the acceptance uncertainty in 
the $\Gpimunu/\Gpienu$ ratio,
we do not require a signal in the muon hodoscope 
to identify \KLpimunu\ decays.

The reconstruction of the $\KLzz$ and  $\KLzzz$ decay modes,
where $\pi^0 \to \gamma\gamma$,
is based on energies and positions of photons measured in the
CsI electromagnetic calorimeter as described in~\cite{Alavi-Harati:2002ye}. 
Exactly four (six) clusters,
each with  a transverse profile consistent with a photon,
are required for $\KLzz$ ($\KLzzz$).
Photons are paired to reconstruct two or three
neutral pions consistent with a single decay vertex. 

All reconstructed decay modes are required to have kaon energy, 
$E_K$, between 40 and 120 GeV, and decay position, $z$,  
between 123 and 158 m from the target.
For the reconstruction of semileptonic and \KLpmz\ decays, 
there is a missing particle ($\nu$ or $\pi^0$), which results in
multiple kaon energy solutions. Each of these solutions is
required to be in the accepted range.

\begin{figure}
  \epsfig{file=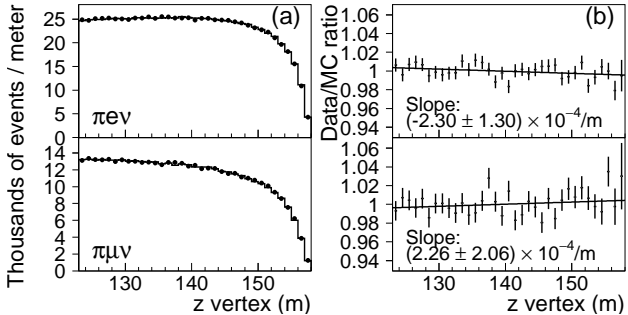, width=\linewidth}
  \caption{
     (a) Comparison of the vacuum beam $z$ distributions for data (dots) 
     and MC (histogram); (b) Data-to-MC ratios with a linear
     fit.
         }
  \label{fig:zvtx_slopes}
\end{figure}

After all event selection requirements and background 
subtraction, 
we have between $10^5$ and $10^6$
events per decay mode.  
The background is 0.7\%\ for \KLzz\ and much smaller for the other 
decay modes.
After correcting each of the ratios for acceptance
differences between numerator and denominator, we find the partial width
ratios given in Table~\ref{tab:ratios}. 
The precision is 1.2\%\ for
$\Gzzz/ \Gpienu$ and about 0.5\%\ for other ratios.
A comparison of data and MC $z$-vertex
distributions for the semileptonic decay modes 
(Fig.~\ref{fig:zvtx_slopes})
demonstrates the quality of the MC simulation
used for the acceptance correction.

\begin{table}
\caption{
    \label{tab:ratios}Measured partial width ratios.
     The first error is statistical and the
     second systematic. The five statistical errors are
     independent; correlations among the systematic
     errors are treated in \cite{ktev_kbr}.
       }
\begin{ruledtabular}
\begin{tabular}{ lc}
Decay Modes       & Partial Width Ratio \\
\hline 
$\Gpimunu/ \Gpienu$  &  $\RPMNvalue \pm \RPMNerrstat \pm \RPMNerrsyst $\\
$\Gzzz/ \Gpienu$  & $\RZZZvalue \pm \RZZZerrstat \pm \RZZZerrsyst $\\ 
$\Gppp/ \Gpienu$  & $\RPPPvalue \pm \RPPPerrstat \pm \RPPPerrsyst $\\ 
$\Gpm/ \Gpienu$  & $(\RPPvalue \pm \RPPerrstat \pm \RPPerrsyst) \times 10^{-3} $\\ 
$\Gzz/ \Gzzz$  & $(\RNEUTvalue \pm \RNEUTerrstat \pm \RNEUTerrsyst ) \times 10^{-3}$\\ 
\end{tabular}\end{ruledtabular}
\end{table}

The five partial width ratios may be combined to determine
the branching fractions shown in Table~\ref{tab:br}~\cite{ktev_kbr}. 
Using the PDG average for the neutral kaon lifetime~\cite{pdg02},
$\tau_L = (5.15 \pm 0.04) \times 10^{-8}$ s, our branching 
fraction measurements correspond to the partial decay widths shown in 
Table~\ref{tab:br}. 

\begin{figure}[ht]
 \epsfig{file=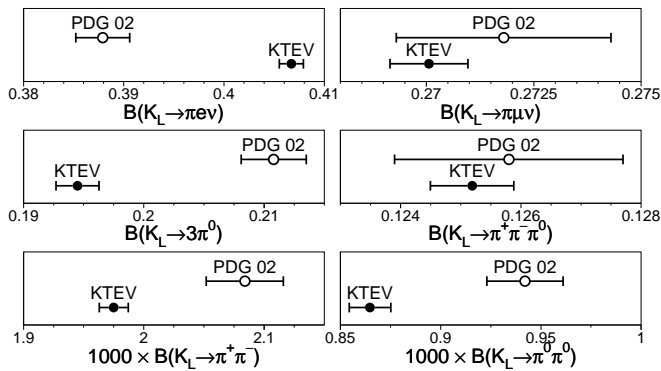, width=\linewidth}
\caption{
    \label{fi:pdg2} 
       $K_L$ branching fractions measured  by KTeV (dots) and from PDG
       fit (open circles).
       }
\end{figure}

Figure~\ref{fi:pdg2}
shows a comparison of the KTeV and PDG values for 
the six branching fractions.
The new KTeV
measurements are on average a factor of two more precise 
than the current world average values, but are not in good agreement
with these averages. 
Compared to the PDG fit~\cite{pdg02},
the KTeV measurement of $B(\KLpienu)$
is higher by 5\%, $B(\KLzzz)$ is lower by 8\%,
$B(\KLpm)$ is lower by 5\%,  and $B(\KLzz)$ is lower by 8\%.
Our measurements of $B(\KLpimunu)$ and $B(\KLpmz)$ are consistent
with the PDG fit.
A detailed comparison between the KTeV measurements 
and previous results is given in~\cite{ktev_kbr}.

Using the measured branching fractions for
$\KLpm$~and~$\KLzz$ together with $\tau_L$, $\tau_S$, $Re(\epe)$, 
and $B(K_S\to\pi\ell\nu)$,
we determine the CP violation parameter
$|\eta_{+-}| = (\etapmv \pm 
\etapmint_{\rm KTeV} \pm \etapmext_{\rm ext})\times 10^{-3}$;
most of the external 
error results from the uncertainty in $\tau_L$.
Our result is 2.6\%\ lower than the PDG average.
A comparison of $|\etapm|$ determinations is given in
\cite{ktev_kbr}.

\begin{table}
\caption{
    \label{tab:br}
     $K_L$ branching fractions and partial widths ($\Gamma_i$).
     Correlations among uncertainties in these measurements
     are given in \cite{ktev_kbr}.
       }
\begin{ruledtabular}
\begin{tabular}{ lcc}
Decay Mode       & Branching fraction & $\Gamma_i$ ($10^7 s^{-1}$) \\
\hline 
\KLpienu & $\BKEvalue \pm \BKEerrtot $ & $\WKEvalue \pm \WKEerrtot $ \\
\KLpimunu & $\BKMvalue \pm \BKMerrtot  $&  $\WKMvalue \pm \WKMerrtot  $ \\
\KLpmz & $\BPMZvalue \pm \BPMZerrtot $ & $\WPMZvalue \pm \WPMZerrtot $  \\
\KLzzz & $\BZZZvalue \pm \BZZZerrtot  $ & $\WZZZvalue \pm \WZZZerrtot  $ \\
\KLpm & $(\BPMvalue \pm \BPMerrtot)\times 10^{-3}  $ 
       &  $(\WPMvalue \pm \WPMerrtot)\times 10^{-3}  $  \\
\KLzz & $(\BZZvalue \pm \BZZerrtot)\times 10^{-3}  $ 
& $(\WZZvalue \pm \WZZerrtot)\times 10^{-3} $ \\
\end{tabular}
\end{ruledtabular}
\end{table}

The \fplust\ form factor is measured
in both semileptonic decay modes; the effect of \fzerot\ is proportional
to the lepton mass, so it is only measured in $\KLpimunu$ decays.
The measured parameters for the semileptonic 
form factors are
$\lplusp = (\lplabv \pm \lplabe)\times 10^{-3}$,
$\lpluspp = (\lplbbv \pm \lplbbe)\times 10^{-3}$, and
$\lzero    = (\lzbv \pm \lzbe )\times 10^{-3}$.
The corresponding phase space integrals are
$\textstyle I_K^{e} = \ikebv \pm \ikebetot $ 
and $\textstyle I_K^{\mu} = \ikmbv \pm \ikmbetot$,
where the quoted errors include an additional uncertainty
related to the form factor parameterization~\cite{ktev_kl3ff}.
Compared to phase space integrals based on PDG form factors,
KTeV's
$I_K^{e}$ and $I_K^{\mu}$ integrals are 1.7\% and 4.2\% lower, 
respectively.
If we fit our data without the $\lpluspp$ term, our
$I_K^{e}$ and $I_K^{\mu}$ integrals are increased by 1\%,
and are
consistent with PDG averages that use only linear terms.

To check the consistency of our branching fraction and form factor
measurements with lepton universality,
we compare $G_F$ for the two decay modes by
taking the ratio of 
Eq.~\ref{eq:vus} for $\KLpimunu$ and $\KLpienu$:
\begin{equation}
     \Big(\frac{G^{\mu}_F}{G^e_F}\Big)^2 =
        \Big[\frac{\textstyle \Rpimunu}{\textstyle \Rpienu} \Big]
             \Big/ 
  \Big(
     \frac{\textstyle 1+ \deltam  }{\textstyle 1 +\deltae}
              \cdot
   \frac{\textstyle I_K^\mu}{\textstyle I_K^e} 
   \Big)
      \label{eq:universal}.
\end{equation} 
Many common uncertainties cancel in this ratio.
The ratio of radiative corrections is calculated to be
$(1+\delta_K^{\mu})/(1+\delta_K^e) = 
\DeltaRat \pm \DeltaRatE$~\cite{Troy}, 
the ratio of the phase space integrals
is $I_K^{\mu}/I_K^e =\ikrbv \pm \ikrbetot$,
and $\Gpimunu/\Gpienu$ is from Table~\ref{tab:br}.
The resulting ratio of couplings squared is 
$(G^{\mu}_F / G^e_F)^2= \LeptUni \pm \LeptUniE$,
consistent with lepton universality.
The same ratio calculated from PDG widths and form factors
is $(G^{\mu}_F / G^e_F)^2= \PDGLeptUni \pm \PDGLeptUniE$.
Note that the 0.5\%\ uncertainty in our universality test is much smaller
than the 5\%\ difference between the KTeV and PDG values of 
$\Gpimunu / \Gpienu$.

The measured partial widths and phase space integrals 
for semileptonic decays can be
combined with theoretical corrections to
calculate $|V_{us}|$ using Eq.~\ref{eq:vus}.
The short-distance radiative correction, $S_{EW} = 1.022$~\cite{sew},
is evaluated with a cutoff at the proton mass.
The long-distance radiative corrections are taken from~\cite{Troy}:
$\delta^e_K = 0.013\pm 0.003$  and 
$\delta^\mu_K = 0.019\pm 0.003$. 
For $f_+(0)$, we use the same value used in the PDG evaluation
of $|V_{us}|$: $f_+(0) = 0.961 \pm 0.008$~\cite{leut-roos}.

The resulting values of $|\vus|$ are
$\vuske \pm \vuskee$ for $K_{e3}$ and
$\vuskm \pm \vuskme$ for $K_{\mu 3}$, where the errors
include an external uncertainty of \vusaveex\ from $f_+(0)$, the
$K_L$ lifetime, and radiative corrections.
Assuming lepton universality, we average the $\KLpienu$ and $\KLpimunu$
results (accounting for correlations):
\begin{equation}
   |\vus| = \vusave \pm \vusavein_{\rm KTeV} \pm \vusaveex_{\rm ext}.
\label{eq:vusresult}
\end{equation}
The KTeV error comes from uncertainties in the KTeV branching fraction
and form factor measurements.

\begin{figure}
\centering
\psfig{figure=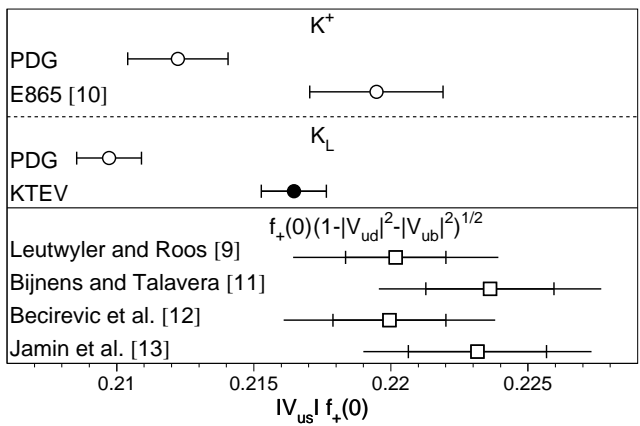,width=\linewidth}
\caption{
Comparison of the KTeV measurement of $|V_{us}|f_+(0)$ 
with Brookhaven E865~\cite{bnl865}, PDG, and also with
determinations of $f_+(0)(1-|V_{ud}|^2-|V_{ub}|^2)^{1/2}$
based on different theoretical calculations of 
$f_+(0)$~\cite{leut-roos, bij, becirevic, jamin}. 
$K^+$ measurements have been divided by $1.022 \pm 0.005$, 
the ratio of $f_+(0)$ for charged
and neutral kaons~\cite{cirigliano}.
For $K_L$ measurements, the uncertainties are mainly from $\tau_L$.
PDG refers to  our evaluation based on PDG  partial
widths and form factors. 
For $f_+(0)(1-|V_{ud}|^2-|V_{ub}|^2)^{1/2}$,
the inner error bars are from $f_+(0)$ uncertainty; the total 
uncertainties include the $|V_{ud}|$ and $|V_{ub}|$ errors.
  }
\label{fig:vus}
\end{figure}

To compare our result with previous charged and neutral kaon  measurements,
we use the product of $|\vus|$ and $f_+(0)$ rather than $|\vus|$
to avoid significant common uncertainties from $f_+(0)$.
Figure~\ref{fig:vus} shows a comparison of our measurement of
\begin{equation}
|\vus|f_+(0) = \vusfave \pm \vusfavee
\end{equation}
with values from the PDG and Brookhaven E865~\cite{bnl865}.  Our value of 
$|V_{us}|f_+(0)$ is inconsistent with previous $K_L$
determinations, but is consistent with $K^+$ results (both E865
and earlier measurements).
The figure also shows 
$f_+(0)(1-|V_{ud}|^2-|V_{ub}|^2)^{1/2}$,
the expectation for $f_+(0)|V_{us}|$ assuming unitarity,
based on $|V_{ud}| = 0.9734 \pm 0.0008$, 
$|V_{ub}| = (3.6 \pm 0.7)\times 10^{-3}$,
and several recent calculations
of $f_+(0)$. Our value of $|V_{us}|$ (Eq.~\ref{eq:vusresult}),
based the Leutwyler and Roos calculation of $f_+(0)$, 
is
consistent with unitarity:
$1-(|V_{ud}|^2+|V_{us}|^2+|V_{ub}|^2)= \deltauni \pm \unite $.
For other calculations of $f_+(0)$,
the consistency with unitarity ranges from 
1 to 1.7 sigma, as shown in
Fig.~\ref{fig:vus}.
Our improved form factor measurements may help to reduce theoretical 
uncertainties in $f_+(0)$~\cite{bij, becirevic, jamin}.

In summary, KTeV has made improved measurements of the six largest $K_L$ 
branching fractions and the semileptonic form factors.  
We use these results to determine
$|\etapm| = (\etapmv \pm \etapme) \times 10^{-3}$ 
and $|\vus|f_+(0) = \vusfave \pm \vusfavee$.
Using $f_+(0) = 0.961 \pm 0.008$~\cite{leut-roos}, we find
$|\vus| = \vusave \pm \vusavee$, consistent with unitarity of
the CKM matrix.

We gratefully acknowledge the support and effort of the Fermilab
staff and the technical staffs of the participating institutions for
their vital contributions. We also thank T. Andre and J. Rosner for 
useful discussions of radiative corrections. This work was supported in 
part by the U.S. 
Department of Energy, The National Science Foundation, and The Ministry of
Education and Science of Japan. 

%


\end{document}